\documentclass[epj,final]{svjour}
\smartqed  
\RequirePackage{graphicx}
\begin{document}
\title{Ultrarelativistic transverse momentum distribution of the Tsallis statistics}
\author{A.S.~Parvan \inst{1,}\inst{2,}\inst{3}
}                     
%
%
\institute{Bogoliubov Laboratory of Theoretical Physics, Joint Institute for Nuclear Research, Dubna, Russia  \and
Department of Theoretical Physics, Horia Hulubei National Institute of Physics and Nuclear Engineering, Bucharest-Magurele, Romania \and
Institute of Applied Physics, Moldova Academy of Sciences, Chisinau, Republic of Moldova}
\date{Received: date / Revised version: date}
%
\abstract{The analytical expressions for the ultrarelativistic transverse momentum distributions of the Tsallis and the Tsallis-$2$ statistics were obtained. We found that the transverse momentum distribution of the Tsallis-factorized statistics, which is now largely used to describe the experimental transverse momentum spectra of hadrons measured in $pp$ collisions at LHC and RHIC energies, in the ultrarelativistic case is not equivalent to the transverse momentum distributions of the Tsallis and the Tsallis-$2$ statistics. However, we revealed that this distribution exactly coincides with the transverse momentum distribution of the Tsallis-$2$ statistics in the zeroth term approximation and is transformed to the transverse momentum distribution of the Tsallis statistics in the zeroth term approximation by changing the parameter $q$ to $1/q_{c}$. We demonstrated analytically on the basis of the ultrarelativistic ideal gas that the Tsallis-factorized statistics is not equivalent to the Tsallis and the Tsallis-$2$ statistics. In the present paper the Tsallis statistics corresponds to the standard expectation values.
\PACS{
      {13.85.-t}{Hadron-induced high- and super-high-energy interactions}   \and
      {13.85.Hd}{Inelastic scattering: many-particle final states} \and
      {24.60.-k}{Statistical theory and fluctuations}
     } 
} 
\titlerunning{Ultrarelativistic transverse momentum distribution}
\authorrunning{A.S.~Parvan}
\maketitle
\section{Introduction}\label{sec1}
At present, many different approximations of the Tsallis statistics are widely used to analyze the LHC and RHIC data on the transverse momentum distributions of hadrons created in the proton-proton and heavy-ion collisions. Both the Tsallis-like momentum distributions~\cite{ALICE09,Atlas1,Cms3} and the classical Maxwell-Boltzmann distributions of the Tsallis- factorized statistics~\cite{Rybczynski14,Cleymans13,Azmi14,Cleymans12a,Cleymans2012,Marques13,Li14,Parvan14,Parvan16a,Biyajima06,Marques15} successfully describe the experimental data of hadrons at high energies. This is due to the fact that in contrast to the Gibbs function the momentum distribution of the Tsallis statistics has a power-law form that allows a proper description of the behavior of hadrons with large transverse momenta.

The Tsallis statistics~\cite{Tsal88,Tsal98} and its modifications (see ref.~\cite{Tsal98} and references therein) by definition are constructed on the basis of the probabilities of microstates of the system instead of the single-particle distribution functions. The many-body distribution function of the Tsallis statistics does not factorize into the product of the single-particle distribution functions because of the power-law form of the probabilities of microstates. This leads to problems in obtaining the single-particle distribution functions of the Tsallis statistics in the explicit form~\cite{Parvan14,Tirnakli00}. In order to avoid the difficulties of analytical and numerical calculations, different approaches were introduced. For instance, there is the factorization (dilute gas) approximation~\cite{Buyukkilic93} of the Tsallis statistics and the Tsallis-factorized statistics~\cite{Lavagno02,Alberico09,Conroy10,Cleymans12a,Cleymans2012}, which allow one to write the single-particle distribution functions of the Tsallis statistics in the explicit form. In the factorization approximation of the Tsallis statistics the many-body distribution function by definition is the product of the single-particle distribution functions. However, the Tsallis-factorized statistics is defined on the basis of the single-particle distribution functions of the ideal gas instead of the probabilities of microstates of the system, and the factorization assumption is contained in it implicitly. Note that through the paper we use the term Tsallis statistics to denote the Tsallis-$1$ statistics introduced in~\cite{Tsal88,Tsal98}. We follow the classification scheme for the Tsallis statistics given in~\cite{Tsal98}. The Tsallis-$1$ statistics uses the standard expectation values. However, the Tsallis-$2$ statistics is based on the generalized expectation values. In ref.~\cite{Tsal98}, the quantities of the Tsallis-$2$ statistics are denoted by the superindex $(2)$ and the quantities of the Tsallis-$1$ statistics are denoted by the superindex $(1)$. The reason why we use the term Tsallis statistics for the standard linear expectation values was clearly explained in~\cite{Parvan2015}.

The main purpose of this study is to find the analytical expression for the transverse momentum distribution of the Tsallis statistics~\cite{Tsal88}, which was developed in~\cite{Parvan06a,Parvan06b,Parvan2015,Parvan2015a}, in the case of the ultrarelativistic Maxwell- Boltzmann particles and demonstrate that the momentum distribution of the Tsallis-factorized statistics~\cite{Cleymans12a,Cleymans2012} in the case of massless particles corresponds to the zeroth term approximation of the Tsallis statistics with transformation of the parameter $q$ to parameter $1/q_{c}$. Note that the thermodynamic self-consistency of the Tsallis statistics~\cite{Tsal88} in the different statistical ensembles was demonstrated in~\cite{Parvan06a,Parvan06b,Parvan2015,Parvan2015a}. The thermodynamic self-consistency of the Tsallis-factorized statistics was proved in~\cite{Cleymans12a,Cleymans2012}.

The structure of the paper is as follows. In Section~\ref{sec2}, we briefly define the main formulas of the Tsallis statistics in the grand canonical ensemble. Section~\ref{sec3} is devoted to the definition of the zeroth term approximation of the Tsallis statistics. In Section~\ref{sec4}, we define the main formulas of the Boltzmann-Gibbs statistics, the Tsallis-$2$ statistics, the zeroth term approximation of the Tsallis-$2$ statistics and the Tsallis-factorized statistics. The main conclusions are summarized in the final section.

\section{Transverse momentum and rapidity distributions of the Tsallis statistics}\label{sec2}
The Tsallis statistics is defined by the generalized entropy with the probabilities $p_{i}$ of the microstates of the system normalized to unity~\cite{Tsal88,Tsal98}
\begin{equation}\label{1}
    S = z \sum\limits_{i}  p_{i} (1-p_{i}^{1/z}),\qquad  \sum\limits_{i} p_{i}=1,
\end{equation}
where $z=1/(q-1)$ and $q\in\mathbf{R}$ is a real parameter taking values $0<q<\infty$. Here and throughout the paper we use the system of natural units $\hbar=c=k_{B}=1$. In the Gibbs limit $|z|\to\infty$ or $q\to 1$, the entropy (\ref{1}) recovers the Boltzmann-Gibbs entropy, $S=-\sum_{i} p_{i} \ln p_{i}$.

In the grand canonical ensemble  $(T,V,z,\mu)$ the probabilities of microstates and the norm equation for the Tsallis statistics are as follows~\cite{Parvan2015}:
\begin{equation}\label{2}
p_{i} = \left[1+\frac{\Lambda-E_{i}+\mu N_{i}}{(z+1)T}\right]^{z}
\end{equation}
and
\begin{equation}\label{3}
\sum\limits_{i} \left[1+\frac{\Lambda-E_{i}+\mu N_{i}}{(z+1)T}\right]^{z}=1,
\end{equation}
where $E_{i}$ and $N_{i}$ are the energy and the number of particles in the $i$-th microscopic state of the system, respectively, and $\Lambda$ is a norm function. In the Gibbs limit $|z|\to\infty$ the probability $p_{i}=\exp[(\Lambda-E_{i}+\mu N_{i})/T]$ and $\Lambda=-T\ln Z$, where $Z=\sum_{i} \exp[-(E_{i}-\mu N_{i})/T]$ is the partition function and $\Lambda$, now, is the thermodynamic potential of the grand canonical ensemble.

Let us consider the Maxwell-Boltzmann statistics of the ultrarelativistic ideal particles in the framework of the Tsallis statistics in the grand canonical ensemble. The thermodynamic potential for the Maxwell-Boltzmann particles in the occupation number representation can be written as~\cite{Parvan2015}
\begin{eqnarray}\label{4}
  && \Omega = \sum\limits_{i}  p_{i} [\Lambda+ T (1-p_{i}^{1/z})]= \frac{z}{z+1} \sum\limits_{\{n_{\vec{p}\sigma}\}} \frac{1}{\prod\limits_{\vec{p}\sigma}n_{\vec{p}\sigma}!}
  \nonumber \\
   && \left[\Lambda+\frac{\sum\limits_{\vec{p}\sigma} n_{\vec{p}\sigma} (\varepsilon_{\vec{p}}-\mu)}{z}\right]  \left[1+\frac{\Lambda- \sum\limits_{\vec{p}\sigma} n_{\vec{p}\sigma} (\varepsilon_{\vec{p}}-\mu)}{(z+1)T}\right]^{z}, \;\;\;\;\;\;
\end{eqnarray}
where $\varepsilon_{\vec{p}}=|\vec{p}|$ is the one-particle energy and $n_{\vec{p}\sigma}=0,1,\ldots,\infty$ are the occupation numbers. Substituting (\ref{2}) into Eq.~(\ref{4}) we obtain
\begin{equation}\label{4a}
  \Omega = \frac{z}{z+1} \left[\Lambda+\frac{E-\mu\langle N\rangle}{z}\right],
\end{equation}
where $E=\sum_{i}  p_{i} E_{i}$ is the energy and $\langle N\rangle=\sum_{i}  p_{i} N_{i}$ is the mean number of particles of the system.

The norm equation (\ref{3}) for the Maxwell-Boltzmann particles of the Tsallis statistics in the grand canonical ensemble takes the form
\begin{equation}\label{5}
  \sum\limits_{\{n_{\vec{p}\sigma}\}} \frac{1}{\prod\limits_{\vec{p}\sigma}n_{\vec{p}\sigma}!}
    \left[1+\frac{\Lambda- \sum\limits_{\vec{p}\sigma} n_{\vec{p}\sigma} (\varepsilon_{\vec{p}}-\mu)}{(z+1)T}\right]^{z} = 1.
\end{equation}
Note that for the negative values of $z$ $(q<1)$ the sum over the occupation numbers in Eq.~(\ref{5}) diverges. This occurs because the power-law function in this sum at large values of the number of particles $N$ and energy $E$ of the system may not suppress a strong growth of the number of microstates of the system corresponding to these values of $N$ and $E$. See Appendix~\ref{App1}. Therefore, in order to extract the physical states of the system at fixed values of $q<1$, the sum should be truncated at large values of $N$. The concrete definition of the cut-off scheme for the negative values of $z$ is given below. In the case when $z$ is positive $(q>1)$ the argument of the power-law function in Eq.~(\ref{5}) may be a negative number at large values of energy $E$. Therefore, for the positive values of $z$ $(q>1)$ we impose the Tsallis cut-off prescription~\cite{TsallCutOff}.

The mean occupation numbers for the Maxwell - Boltzmann particles of the Tsallis statistics can be written as
\begin{equation}\label{6}
  \langle n_{\vec{p}\sigma} \rangle = \sum\limits_{\{n_{\vec{p}\sigma}\}} n_{\vec{p}\sigma} \frac{1}{\prod\limits_{\vec{p}\sigma}n_{\vec{p}\sigma}!}
    \left[1+\frac{\Lambda- \sum\limits_{\vec{p}\sigma} n_{\vec{p}\sigma} (\varepsilon_{\vec{p}}-\mu)}{(z+1)T}\right]^{z}.
\end{equation}
The mean number of particles of the system in the grand canonical ensemble is given by
\begin{eqnarray}\label{7}
    \langle N\rangle &=& -\left(\frac{\partial \Omega}{\partial \mu}\right)_{TVz} = \sum\limits_{\vec{p}\sigma}  \langle n_{\vec{p}\sigma} \rangle \nonumber \\
    &=& \sum\limits_{\{n_{\vec{p}\sigma}\}} \frac{ \left(\sum\limits_{\vec{p}\sigma} n_{\vec{p}\sigma}\right)}{\prod\limits_{\vec{p}\sigma}n_{\vec{p}\sigma}!}
    \left[1+\frac{\Lambda- \sum\limits_{\vec{p}\sigma} n_{\vec{p}\sigma} (\varepsilon_{\vec{p}}-\mu)}{(z+1)T}\right]^{z}. \;\;\;
\end{eqnarray}
The energy of the system in the grand canonical ensemble is
\begin{eqnarray}\label{7a}
    E &=& -T^{2}\frac{\partial}{\partial T}\left(\frac{\Omega}{T}\right)_{Vz\mu} + \mu \langle N\rangle = \sum\limits_{\vec{p}\sigma}  \langle n_{\vec{p}\sigma} \rangle \varepsilon_{\vec{p}} \nonumber \\
    &=& \sum\limits_{\{n_{\vec{p}\sigma}\}} \frac{ \left(\sum\limits_{\vec{p}\sigma} n_{\vec{p}\sigma}\varepsilon_{\vec{p}}\right)}{\prod\limits_{\vec{p}\sigma}n_{\vec{p}\sigma}!}
    \left[1+\frac{\Lambda- \sum\limits_{\vec{p}\sigma} n_{\vec{p}\sigma} (\varepsilon_{\vec{p}}-\mu)}{(z+1)T}\right]^{z}. \;\;\;
\end{eqnarray}
The entropy of the system $S$ for the Maxwell-Boltzmann particles of the Tsallis statistics can be written as
\begin{eqnarray}\label{8}
   S &=&  -\left(\frac{\partial \Omega}{\partial T}\right)_{Vz\mu} =  -z \sum\limits_{\{n_{\vec{p}\sigma}\}} \frac{\Lambda- \sum\limits_{\vec{p}\sigma} n_{\vec{p}\sigma} (\varepsilon_{\vec{p}}-\mu)}{(z+1)T} \nonumber \\
    && \frac{1}{\prod\limits_{\vec{p}\sigma}n_{\vec{p}\sigma}!}  \left[1+\frac{\Lambda- \sum\limits_{\vec{p}\sigma} n_{\vec{p}\sigma} (\varepsilon_{\vec{p}}-\mu)}{(z+1)T}\right]^{z}. \qquad
\end{eqnarray}
Then the entropy (\ref{8}) and the thermodynamic potential (\ref{4}) satisfy the relations
\begin{equation}\label{8a}
S = -\frac{z}{z+1}\frac{\Lambda-E+\mu \langle N\rangle}{T}=z\frac{\Omega-\Lambda}{T}
\end{equation}
and
\begin{equation}\label{8b}
    S =-\frac{\Omega-E+\mu \langle N\rangle}{T},
\end{equation}
where the last equation is the Legendre transform.

The transverse momentum and rapidity distributions of particles can be defined as
\begin{equation}\label{9}
  \frac{d^{2}N}{dp_{T}dy} = \frac{V}{h^{3}} \sum\limits_{\sigma} \int\limits_{0}^{2\pi} d\varphi p_{T} \varepsilon_{\vec{p}} \  \langle n_{\vec{p}\sigma}\rangle
\end{equation}
and
\begin{equation}\label{10}
  \frac{dN}{dy} = \frac{V}{h^{3}} \sum\limits_{\sigma} \int\limits_{0}^{2\pi} d\varphi \int\limits_{0}^{\infty} dp_{T} p_{T} \varepsilon_{\vec{p}} \  \langle n_{\vec{p}\sigma}\rangle,
\end{equation}
where $\varepsilon_{\vec{p}}=p_{T} \cosh y$ for the ultrarelativistic particles, $p_{T}$ and $y$ are the transverse momentum and rapidity, respectively, and $\langle n_{\vec{p}\sigma}\rangle$ are the mean occupation numbers given by Eq.~(\ref{6}).

Let us calculate the sums in Eqs.~(\ref{4}) and (\ref{5})-(\ref{8}). To obtain the exact results we use the integral representations for the Gamma-function~\cite{Abramowitz,Prato}
\begin{eqnarray}\label{11}
  x^{-y} &=& \frac{1}{\Gamma(y)} \int\limits_{0}^{\infty}  t^{y-1} e^{-tx}  dt , \;\;\;\; Re(x)>0, Re(y)>0, \\ \label{12}
   x^{y-1} &=& \Gamma(y) \frac{i}{2\pi} \oint\limits_{C} (-t)^{-y} e^{-tx}  dt , Re(x)>0, |y|<\infty. \;\;\;\;\;\;
\end{eqnarray}
Substituting Eq.~(\ref{11}) for $z<-1$ and Eq.~(\ref{12}) for $z>0$ into (\ref{5}) and using the partition function for the Maxwell-Boltzmann ultrarelativistic ideal gas in the Boltzmann-Gibbs statistics in the grand canonical ensemble
\begin{equation}\label{13}
  Z=\exp\left(\frac{gVT^{3}}{\pi^{2}} \ e^{\mu/T}\right),
\end{equation}
we obtain
\begin{equation}\label{14}
   \sum\limits_{N=0}^{N_{0}} \frac{\tilde{\omega}^{N}}{N!} h_{0}(0) \left[1+\frac{\Lambda+\mu N}{(z+1)T}\right]^{z+3N}=1,
\end{equation}
where
\begin{equation}\label{15}
   h_{\eta}(\xi) = \frac{(-z-1)^{3(N+\eta)}  \Gamma(-z-\xi-3(N+\eta))}{\Gamma(-z-\xi)}, \;\;  z<-1
\end{equation}
and
\begin{equation}\label{16}
   h_{\eta}(\xi) = \frac{(z+1)^{3(N+\eta)}\Gamma(z+1+\xi)}{\Gamma(z+1+\xi+3(N+\eta))}, \quad  z>0.
\end{equation}
Here, $\tilde{\omega}=gVT^{3}/\pi^{2}$ and $g$ is the spin degeneracy factor. For the negative values of $z$ $(q<1)$ the upper bound of summation $N_{0}$ in Eq.~(\ref{14}) is fixed from the condition $N<-z/3$ and the inflection point of the logarithm of the function
\begin{equation}\label{17}
  \phi(N) = \frac{\tilde{\omega}^{N}}{N!} h_{0}(0) \left[1+\frac{\Lambda+\mu N}{(z+1)T}\right]^{z+3N}.
\end{equation}
Thus, $N=N_{0}$ is the solution of the equation
\begin{equation}\label{17b}
  \frac{\partial^{2}\ln \phi(N)}{\partial N^{2}}=0.
\end{equation}
This cut-off represents our regularization prescription for the divergencies appearing at negative values of $z$. See Appendix~\ref{App1}. For the positive values of $z$ $(q>1)$ the upper bound of summation $N_{0}$ in Eq.~(\ref{14}) should be determined from the conditions $1+(\Lambda+\mu N)/((z+1)T)>0$ and $z+1+3N<\infty$. Note that after solving Eq.~(\ref{14}) the norm function $\Lambda$ becomes a function of the variables of state $(T,V,z,\mu)$. In the Gibbs limit $|z|\to\infty$ the norm function $\Lambda$ from Eq.~(\ref{14}) recovers its Boltzmann-Gibbs value
\begin{equation}\label{17a}
  \Lambda=-T\ln Z=-\frac{gVT^{4}}{\pi^{2}} \ e^{\mu/T}
\end{equation}
and it is the thermodynamic potential of the grand canonical ensemble.

It is worth mentioning that in the case of $q<1$ the cut-off parameter $N_{0}$ in Eq.~(\ref{14}) can be found also from the local minimum of the function $\ln\phi(N)$. However, in this paper, we fix the cut-off parameter $N_{0}$ only from the inflection point of the function $\ln\phi(N)$.

Substituting Eqs.~(\ref{11}), (\ref{12}) into (\ref{6}) and using the partition function (\ref{13}) and the mean occupation numbers of the Maxwell-Boltzmann ultrarelativistic ideal gas in the Boltzmann-Gibbs statistics
\begin{equation}\label{19}
  \langle n_{\vec{p}\sigma}\rangle=e^{-\frac{\varepsilon_{\vec{p}}-\mu}{T}},
\end{equation}
we obtain
\begin{equation}\label{20}
  \langle n_{\vec{p}\sigma}\rangle =  \sum\limits_{N=0}^{N_{0}} \frac{\tilde{\omega}^{N}}{N!} h_{0}(0) \left[1+\frac{\Lambda-\varepsilon_{\vec{p}}+\mu (N+1)}{(z+1)T}\right]^{z+3N},
\end{equation}
where $\Lambda$ is calculated from Eqs.~(\ref{14})--(\ref{16}). For negative $z<-1$ the upper bound of summation $N_{0}$ is the same as in Eq.~(\ref{14}). However, for positive values of $z$ $(q>1)$ the upper limit $N_{0}$ should be determined from the conditions $1+(\Lambda-\varepsilon_{\vec{p}}+\mu (N+1))/((z+1)T)>0$ and $z+1+3N<\infty$. Thus, at fixed values of $z>0$ the one-particle energies are restricted by the condition $\varepsilon_{\vec{p}}<\Lambda +\mu+T (z+1)$. In the Gibbs limit $|z|\to\infty$ the mean occupation numbers (\ref{20}) resemble their Boltzmann-Gibbs values (\ref{19}).

\begin{figure*}
\includegraphics[width=1.0\textwidth]{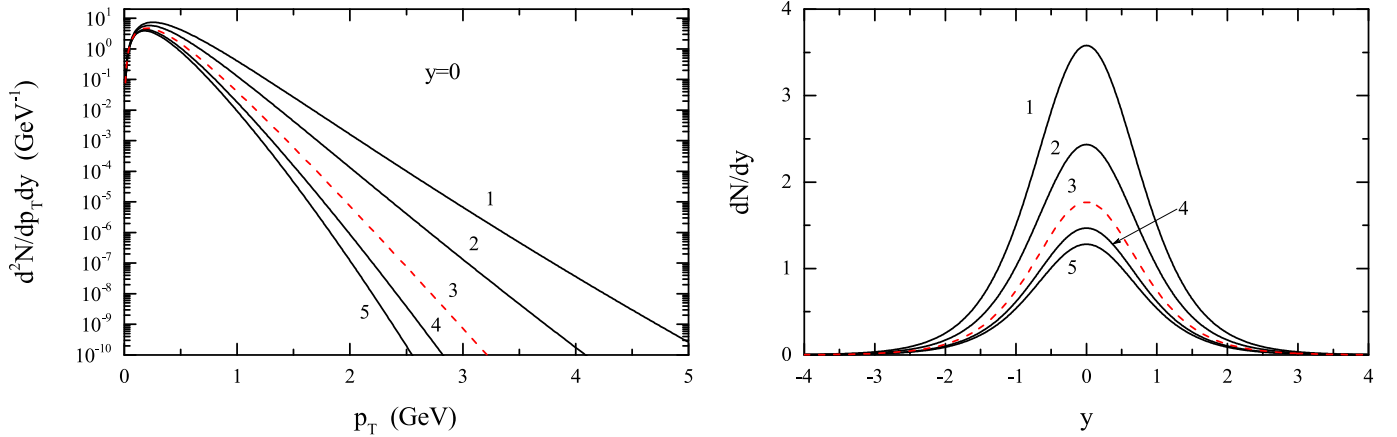}
\caption{(Color online) Transverse momentum and rapidity distributions for the ultrarelativistic $\pi^{-}$ pions in the Tsallis statistics (solid lines) and the Boltzmann-Gibbs statistics (dashed lines) at the temperature $T=100$ MeV, radius $R=4$ fm, $\mu=0$ and different values of the parameter $q$. The lines $1,2,3,4$ and $5$ are calculations for $q=0.99,0.995,1.0,1.005$ and $1.01$, respectively. } \label{fig1}
\end{figure*}

Substituting Eqs.~(\ref{11}), (\ref{12}) into (\ref{7}) and using the partition function (\ref{13}) and the mean number of particles for the Maxwell-Boltzmann ultrarelativistic ideal gas in the Boltzmann-Gibbs statistics
\begin{equation}\label{21}
  \langle N \rangle=\frac{gVT^{3}}{\pi^{2}} \ e^{\mu/T},
\end{equation}
we have
\begin{equation}\label{22}
  \langle N \rangle = \sum\limits_{N=0}^{N_{0}} \frac{\tilde{\omega}^{N+1}}{N!} h_{1}(0) \left[1+\frac{\Lambda+\mu (N+1)}{(z+1)T}\right]^{z+3(N+1)},
\end{equation}
where the upper bound of summation $N_{0}$ for $z<-1$ is the same as in Eq.~(\ref{14}). For $z>0$, $N_{0}$ is determined from the conditions $1+(\Lambda+\mu (N+1))/((z+1)T)>0$ and $z+1+3(N+1)<\infty$. In the Gibbs limit $|z|\to\infty$ the mean number of particles (\ref{22}) resembles its Boltzmann-Gibbs value (\ref{21}).

Substituting Eqs.~(\ref{11}), (\ref{12}) into (\ref{7a}) and using the partition function (\ref{13}) and the energy for the Maxwell-Boltzmann ultrarelativistic ideal gas in the Boltzmann-Gibbs statistics
\begin{equation}\label{21a}
  E = \frac{3gVT^{4}}{\pi^{2}} \ e^{\mu/T},
\end{equation} we have
\begin{equation}\label{22a}
  E = 3T\sum\limits_{N=0}^{N_{0}} \frac{\tilde{\omega}^{N+1}}{N!} h_{1}(1) \left[1+\frac{\Lambda+\mu (N+1)}{(z+1)T}\right]^{z+1+3(N+1)},
\end{equation}
where the upper bound of summation $N_{0}$ for $z<-1$ is the same as in Eq.~(\ref{14}). For $z>0$, $N_{0}$ is determined from the conditions $1+(\Lambda+\mu (N+1))/((z+1)T)>0$ and $z+2+3(N+1)<\infty$. In the Gibbs limit $|z|\to\infty$ the energy (\ref{22a}) resembles its Boltzmann-Gibbs value (\ref{21a}).

Using Eqs.~(\ref{4}), (\ref{11})--(\ref{13}), (\ref{21}) and (\ref{21a}) we obtain the thermodynamic potential for the ultrarelativistic ideal gas in the grand canonical ensemble as
\begin{eqnarray}\label{18}
  \Omega &=& \frac{z}{z+1}\Lambda+ \frac{1}{z+1} T \left[3-\frac{\mu}{T}+\frac{3\Lambda}{(z+1)T} \right] \nonumber \\
   && \sum\limits_{N=0}^{N_{0}} \frac{\tilde{\omega}^{N+1}}{N!} h_{1}(1) \left[1+\frac{\Lambda+\mu (N+1)}{(z+1)T}\right]^{z+3(N+1)}, \qquad
\end{eqnarray}
where the upper bound of summation $N_{0}$ is the same as in Eq.~(\ref{14}) for $z<-1$ and is the same as in Eq.~(\ref{22a}) for $z>0$. In the Gibbs limit $|z|\to\infty$ the thermodynamic potential (\ref{18}) recovers the Boltzmann-Gibbs thermodynamic potential (\ref{17a})
\begin{equation}\label{18a}
  \Omega=\Lambda=-\frac{gVT^{4}}{\pi^{2}} \ e^{\mu/T}.
\end{equation}

Using Eqs.~(\ref{8}), (\ref{11})--(\ref{13}), (\ref{21}) and (\ref{21a}) we obtain the entropy for the ultrarelativistic ideal gas in the grand canonical ensemble as
\begin{eqnarray}\label{23}
  S &=&-\frac{z\Lambda}{(z+1)T}+ \frac{z}{z+1} \left[3-\frac{\mu}{T}+\frac{3\Lambda}{(z+1)T} \right] \nonumber \\
  &&\sum\limits_{N=0}^{N_{0}} \frac{\tilde{\omega}^{N+1}}{N!} h_{1}(1) \left[1+\frac{\Lambda+\mu (N+1)}{(z+1)T}\right]^{z+3(N+1)}, \qquad
\end{eqnarray}
where the upper bound of summation $N_{0}$ is the same as in Eq.~(\ref{14}) for $z<-1$ and is the same as in Eq.~(\ref{22a}) for $z>0$. In the Gibbs limit $|z|\to\infty$ the entropy (\ref{23}) recovers the Boltzmann-Gibbs entropy for the ultrarelativistic ideal gas in the grand canonical ensemble
\begin{equation}\label{23a}
  S=\left(4-\frac{\mu}{T}\right) \frac{gVT^{3}}{\pi^{2}} \ e^{\mu/T}.
\end{equation}

Substituting Eq.~(\ref{20}) into Eq.~(\ref{9}) we obtain the transverse momentum distribution of the Maxwell - Boltzmann ultrarelativistic particles in the grand canonical ensemble of the Tsallis statistics as
\begin{eqnarray}\label{24}
  \frac{d^{2}N}{dp_{T}dy} &=& \frac{gV}{(2\pi)^{2}} p_{T}^{2} \cosh y  \   \sum\limits_{N=0}^{N_{0}} \frac{\tilde{\omega}^{N}}{N!} h_{0}(0) \nonumber \\
   &\times&\left[1+\frac{\Lambda-p_{T} \cosh y +\mu (N+1)}{(z+1)T}\right]^{z+3N},
\end{eqnarray}
where the upper bound of summation $N_{0}$ is the same as in Eq.~(\ref{14}) for $z<-1$. However, for $z>0$ the upper limit $N_{0}$ is fixed from the conditions $1+(\Lambda-p_{T} \cosh y+\mu (N+1))/((z+1)T)>0$ and $z+1+3N<\infty$ and the transverse momentum of particles is restricted by its maximal value $p_{T}^{max}=(\Lambda+\mu +T(z+1))/\cosh y$. In the Gibbs limit $|z|\to \infty$ Eq.~(\ref{24}) recovers the Maxwell - Boltzmann transverse momentum distribution of the Boltzmann - Gibbs statistics
\begin{equation}\label{24a}
  \frac{d^{2}N}{dp_{T}dy} = \frac{gV}{(2\pi)^{2}} p_{T}^{2} \cosh y  \  e^{-\frac{p_{T} \cosh y -\mu}{T}}.
\end{equation}

Integrating Eq.~(\ref{24}) with respect to $p_{T}$ from $0$ to $\infty$ results in the rapidity distribution of the Tsallis statistics for the Maxwell - Boltzmann ultrarelativistic particles in the grand canonical ensemble
\begin{equation}\label{25}
  \frac{dN}{dy} = \frac{\langle N \rangle}{2\cosh^{2}y},
\end{equation}
where $\langle N \rangle$ is the mean number of particles given in Eq.~(\ref{22}). In the Gibbs limit $|z|\to\infty$ the mean number of particles $\langle N \rangle$ in Eq.~(\ref{25}) is given by Eq.~(\ref{21}).

Figure~\ref{fig1} shows the transverse momentum distribution (left panel) at rapidity $y=0$ and the rapidity distribution (right panel) for the ultrarelativistic $\pi^{-}$ pions in the Tsallis and Boltzmann-Gibbs statistics at the temperature $T=100$ MeV, radius $R=4$ fm, $\mu=0$ and different values of the parameter $q$. With decrease of the values of the parameter $q$ in comparison with the unity the Tsallis transverse momentum distribution considerably deviates from the Boltzmann-Gibbs transverse momentum distribution in the direction of larger values of $p_{T}$. On the contrary, with increase of the values of the parameter $q$ in comparison with the unity the Tsallis transverse momentum distribution deviates from the Boltzmann-Gibbs transverse momentum distribution in the direction of smaller values of $p_{T}$. The values of the parameter $q>1$ suppress the appearance of particles with large values of the transverse momentum. However, the values of the parameter $q<1$ contribute to the creation of particles with a large transverse momentum. The Boltzmann-Gibbs transverse momentum distribution rapidly decreases with $p_{T}$ and does not describe the experimental data of hadrons created in the high-energy collisions. These distributions of hadrons in the high-energy collisions may be satisfactorily described only by the values of the Tsallis parameter $q<1$.

Both the Tsallis and Boltzmann-Gibbs rapidity distributions have the form (\ref{25}) and are proportional to the mean number of particles $\langle N \rangle$. With decrease of the values of the parameter $q$ in comparison with the unity the Tsallis rapidity distribution increases and becomes broader in comparison with the Boltzmann-Gibbs distribution at $q=1$ and the mean number of particles also grows. See the right panel of Fig.~\ref{fig1}. On the contrary, with increase of the values of the parameter $q$ in comparison with the unity the Tsallis rapidity distribution decreases and becomes narrower in comparison with the Boltzmann-Gibbs distribution and the mean number of particles also decreases. In general, the mean number of particles of the system decreases with $q$. For instance, for $q=0.99,0.995,1.0,1.005$ and $1.01$ we have the mean number of particles $\langle N \rangle=7.2,4.9,3.5,2.9$ and $2.6$, respectively.

\section{The zeroth term approximation for $q<1$}\label{sec3}
\begin{figure}
\includegraphics[width=0.46\textwidth]{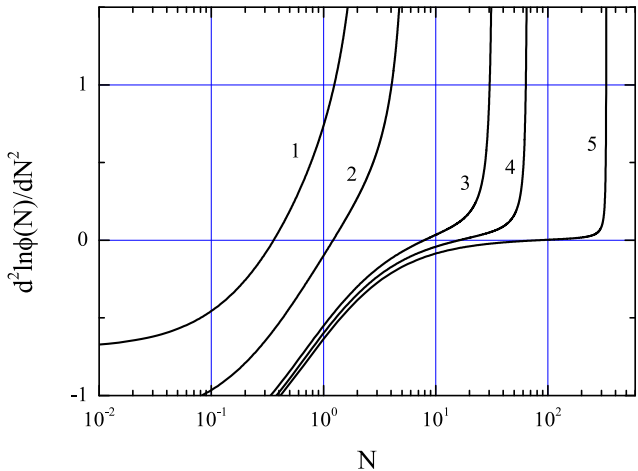}
\caption{(Color online) The second derivative of the logarithm of the function (\ref{17}) with respect to $N$ as a function of $N$ at the temperature $T=100$ MeV, radius $R=4$ fm, $\mu=0$ and different values of the parameter $q$. The lines $1,2,3,4$ and $5$ are calculations for $q=0.9,0.95,0.99,0.995$ and $0.999$, respectively. } \label{fig2}
\end{figure}

The value of the upper bound of summation $N_{0}$ in Eq.~(\ref{14}) decreases with deviation of the value of $q$ from the unity in the range $q<1$. At large deviations of $q$ from the unity the upper bound of summation vanishes, $N_{0}=0$. See Fig.~\ref{fig2}, which represents the second derivative of the logarithm of the function (\ref{17}) with respect to $N$ as a function of $N$ at the temperature $T=100$ MeV, radius $R=4$ fm, $\mu=0$ and different values of the parameter $q$. For $q=0.9,0.95,0.99,0.995$ and $0.999$ we have $N_{0}=0,1,7,16$ and $82$, respectively. However, the condition that the energy (\ref{22a}) of the Tsallis statistics has a finite value leads to the constraint that $q>3/4$.

Let us define the zeroth term approximation for the Tsallis statistics at $q<1$ by introducing the cut-off prescription $N_{0}=0$. Solving Eq.~(\ref{14}) under the condition $N_{0}=0$, we obtain the norm function $\Lambda=0$. Substituting the value of $\Lambda$ into Eq.~(\ref{20}) and considering the upper bound $N_{0}=0$ we obtain the mean occupation numbers as
\begin{equation}\label{26}
  \langle n_{\vec{p}\sigma}\rangle = \left[1-\frac{q-1}{q}\frac{\varepsilon_{\vec{p}}-\mu}{T}\right]^{\frac{1}{q-1}}.
\end{equation}
In the Gibbs limit $q\to 1$, we have Eq.~(\ref{19}), i.e., $\langle n_{\vec{p}\sigma}\rangle=\exp(-\frac{\varepsilon_{\vec{p}}-\mu}{T})$. The Tsallis mean number of particles (\ref{22}) at $N_{0}=0$ takes the form
\begin{equation}\label{27}
  \langle N \rangle = \frac{gVT^{3}}{\pi^{2}}
  \frac{\left(\frac{q}{1-q}\right)^{3}\Gamma\left(\frac{1}{1-q}-3\right)}{\Gamma\left(\frac{1}{1-q}\right)} \left[1+\frac{q-1}{q}\frac{\mu}{T}\right]^{\frac{1}{q-1}+3}.
\end{equation}
In the Gibbs limit $q\to 1$ Eq.~(\ref{27}) recovers the mean number of particles (\ref{21}) of the Boltzmann - Gibbs statistics, $\langle N \rangle=(gVT^{3}/\pi^{2}) e^{\mu/T}$.
The Tsallis energy (\ref{22a}) in the zeroth term approximation can be written as
\begin{equation}\label{27a}
  E = \frac{3gVT^{4}}{\pi^{2}}
  \frac{\left(\frac{q}{1-q}\right)^{3}\Gamma\left(\frac{q}{1-q}-3\right)}{\Gamma\left(\frac{q}{1-q}\right)} \left[1+\frac{q-1}{q}\frac{\mu}{T}\right]^{\frac{q}{q-1}+3}.
\end{equation}
In the Gibbs limit $q\to 1$ Eq.~(\ref{27a}) recovers the energy (\ref{21a}) of the Boltzmann - Gibbs statistics, $E=(3gVT^{4}/\pi^{2}) e^{\mu/T}$. The energy (\ref{27a}) for the Tsallis statistics in the zeroth term approximation also leads to the condition $q>3/4$.

In the zeroth term approximation the thermodynamic potential (\ref{18}) is rewritten as
\begin{eqnarray}\label{27b}
  \Omega &=& -\frac{gVT^{4}}{\pi^{2}} \frac{q-1}{q} \left(3-\frac{\mu}{T}\right) \nonumber \\
  && \frac{\left(\frac{q}{1-q}\right)^{3}\Gamma\left(\frac{q}{1-q}-3\right)}{\Gamma\left(\frac{q}{1-q}\right)} \left[1+\frac{q-1}{q}\frac{\mu}{T}\right]^{\frac{1}{q-1}+3}. \quad
\end{eqnarray}
In the Gibbs limit $q\to 1$ the thermodynamic potential (\ref{27b}) vanishes, $\Omega=0$ and it differs from Eq.~(\ref{18a}). The entropy (\ref{23}) in the zeroth term approximation can be rewritten as
\begin{eqnarray}\label{27c}
  S &=& \frac{gVT^{3}}{\pi^{2}} \left(3-\frac{\mu}{T}\right) \nonumber \\
  && \frac{\frac{1}{q}\left(\frac{q}{1-q}\right)^{3}\Gamma\left(\frac{q}{1-q}-3\right)}{\Gamma\left(\frac{q}{1-q}\right)} \left[1+\frac{q-1}{q}\frac{\mu}{T}\right]^{\frac{1}{q-1}+3}. \quad
\end{eqnarray}
Note that Eqs.~(\ref{27})--(\ref{27c}) satisfy relations (\ref{8a}) and (\ref{8b}). Using Eqs.~(\ref{26}) and (\ref{27c}) we obtain
\begin{equation}\label{27d}
  S=-\sum\limits_{\vec{p}\sigma} \langle n_{\vec{p}\sigma}\rangle^{q} \ln_{q}\langle n_{\vec{p}\sigma}\rangle,
\end{equation}
where $\ln_{q}(x)=(x^{1-q}-1)/(1-q)$. In the Gibbs limit $q\to 1$ the entropy (\ref{27c}) takes the form
\begin{equation}\label{27e}
  S=\left(3-\frac{\mu}{T}\right) \frac{gVT^{3}}{\pi^{2}} \ e^{\mu/T}.
\end{equation}
Equation (\ref{27e}) does not recover the entropy of the Boltzmann - Gibbs statistics (\ref{23a}). Thus, in the Gibbs limit $q\to 1$ the thermodynamic potential and entropy of the Tsallis statistics in the zeroth term approximation do not resemble their corresponding relations of the Boltzmann - Gibbs statistics because the norm function $\Lambda$ in this approximation is equal to zero and the zeroth term approximation in general is reliable only for large deviations of the parameter $q$ from the unity at $q<1$. Hence, the zeroth term approximation of the Tsallis statistics is not consistent.

In the zeroth term approximation the Tsallis transverse momentum distribution (\ref{24}) can be rewritten as
\begin{equation}\label{28}
  \frac{d^{2}N}{dp_{T}dy} = \frac{gV p_{T}^{2} \cosh y}{(2\pi)^{2}}  \left[1-\frac{q-1}{q}\frac{p_{T} \cosh y -\mu}{T}\right]^{\frac{1}{q-1}}.
\end{equation}
In the Gibbs limit $q\to 1$ Eq.~(\ref{28}) resembles the Maxwell - Boltzmann transverse momentum distribution of the Boltzmann - Gibbs statistics (\ref{24a}). Note that the zeroth term approximation is valid only at large deviations of $q$ from unity at $q<1$.

The rapidity distribution $dN/dy$ in the zeroth term approximation is calculated from Eq.~(\ref{25}) with $\langle N \rangle $ given by Eq.~(\ref{27}).

The transverse momentum distribution used in Ref.~\cite{Cleymans2012} corresponds to identifying
\begin{equation}\label{29}
  q\to \frac{1}{q_{c}}.
\end{equation}
After this substitution Eq.~(\ref{28}) becomes
\begin{equation}\label{30}
  \frac{d^{2}N}{dp_{T}dy} = \frac{gVp_{T}^{2} \cosh y}{(2\pi)^{2}}  \left[1+(q_{c}-1)\frac{p_{T} \cosh y -\mu}{T}\right]^{\frac{q_{c}}{1-q_{c}}}.
\end{equation}
Here the parameter $q$ from Ref.~\cite{Cleymans2012} was denoted as $q_{c}$. Thus, the ultrarelativistic transverse momentum distribution of the Tsallis-factorized statistics (\ref{30}), which was defined in Ref.~\cite{Cleymans2012}, is obtained from the ultrarelativistic transverse momentum distribution (\ref{28}) of the Tsallis statistics in the zeroth term approximation by the transformation $q\to 1/q_{c}$. The presence of the different parameters ($q$ and $q_{c}$) in these two distributions is due to the fact that the Tsallis-factorized statistics is defined on the basis of the generalized expectation values, however, the Tsallis statistics is defined on the basis of the standard expectation values. However, the reducibility between these two transverse momentum distributions can be explained by the fact that the Tsallis-factorized statistics is defined on the basis of the single-particle distribution functions and the zeroth term approximation of the Tsallis statistics is defined on the basis of the quantities in which the higher-order terms ($N\geq 1$) were neglected.

It is worth mentioning that the ultrarelativistic transverse momentum distribution of the Tsallis- factorized statistics (\ref{30}) does not recover the ultrarelativistic transverse momentum distribution (\ref{24}) of the Tsallis statistics because the transverse momentum distribution of the Tsallis-factorized statistics corresponds only to the zero term ($N=0$) in the sum (\ref{24}) when the parameter $q$ is changed by the parameter $1/q_{c}$ and $\Lambda$ is found from Eq.~(\ref{14}) at the condition $N_{0}=0$. The higher-order terms ($N\geq 1$) in the sum (\ref{24}) appear due to the fact that the Tsallis statistics is defined on the basis of the probabilities of microstates of the system. However, the absence of the higher-order terms ($N\geq 1$) in the transverse momentum distribution of the Tsallis-factorized statistics (\ref{30}) is due to the fact that the Tsallis-factorized statistics is defined on the basis of the single-particle distribution functions of the ideal gas, which correspond to the zeroth term approximation of the Tsallis statistics at $N_{0}=0$. Hence, the Tsallis-factorized statistics does not correspond to the Tsallis statistics because of the different definitions of the probability of states as well as the different definitions of the expectation values.

Let us examine in more detail these properties by comparing the Tsallis-factorized statistics with the Tsallis-$2$ statistics which both are defined on the basis of the generalized expectation values.

Note that the ultrarelativistic transverse momentum distribution (\ref{24}) of the Tsallis statistics and the ultrarelativistic transverse momentum distribution of the Tsallis- factorized statistics (\ref{30}) were numerically compared and applied to describe the experimental data on the transverse momentum distributions of hadrons at high energies in~\cite{Parvan16}. In ref.~\cite{Parvan16}, the numerical results for the ultrarelativistic transverse momentum distribution (\ref{24}) of the Tsallis statistics were obtained by the cut-off parameter $N_{0}$, which was derived only from the inflection point of the function $\ln\phi(N)$. The case when $N_{0}$ is derived from the local minimum of the function $\ln\phi(N)$ was not studied yet.

\section{Nonequivalence of the Tsallis-factorized statistics with the Tsallis statistics}\label{sec4}
\subsection{Boltzmann-Gibbs statistics}
In the statistical mechanics the thermodynamic potential of the grand canonical ensemble $\Omega$ is obtained from the thermodynamic potential of the fundamental ensemble by the Legendre transform (\ref{8b}), i.e. $\Omega=E-TS-\mu\langle N\rangle$. Then the entropy and the thermodynamic potential of the Boltzmann-Gibbs statistics can be written in the general form as
\begin{equation}\label{1c}
  S = -\sum\limits_{i} p_{i} \ln p_{i},
\end{equation}
and
\begin{equation}\label{2c}
  \Omega = T\sum\limits_{i} p_{i}\left[\ln p_{i} + \frac{E_{i}-\mu N_{i}}{T}\right],
\end{equation}
where $E_{i}$ and $N_{i}$ are the energy and the number of particles, respectively, in the $i$-th microscopic state of the system and $p_{i}$ is the probability of the $i$-th microstate (the many-body distribution function). The unknown probabilities $\{p_{i}\}$ can be obtained from the constrained local extrema of the thermodynamic potential (\ref{2c}) by the method of the Lagrange multipliers~\cite{Parvan2015}
\begin{eqnarray}\label{3c}
 \Phi &=& \Omega - \lambda \varphi, \\ \label{4c}
 \varphi &=& \sum\limits_{i} p_{i} - 1 = 0, \\ \label{5c}
  \frac{\partial \Phi}{\partial p_{i}} &=& 0,
\end{eqnarray}
where $\lambda$ is the Lagrange multiplier. Then the many-body distribution function for the Boltzmann-Gibbs statistics in the grand canonical ensemble is
\begin{eqnarray}\label{6c}
p_{i} &=& \frac{1}{Z} \ e^{-\frac{1}{T}(E_{i}-\mu N_{i})}, \\ \label{7c}
    Z &=& \sum\limits_{i} e^{-\frac{1}{T}(E_{i}-\mu N_{i})},
\end{eqnarray}
where $Z$ is the partition function.

Let us consider the Maxwell-Boltzmann ideal gas. Then the single-particle distribution function and the partition function of the ideal gas in the Boltzmann-Gibbs statistics are derived from Eqs.~(\ref{6c}) and (\ref{7c}) as
\begin{eqnarray}\label{8c}
  \langle n_{\vec{p}\sigma} \rangle &=& \frac{1}{Z}\sum\limits_{\{n_{\vec{p}\sigma}\}} n_{\vec{p}\sigma} \frac{1}{\prod\limits_{\vec{p}\sigma}n_{\vec{p}\sigma}!} \
  e^{-\frac{1}{T} \sum\limits_{\vec{p}\sigma} n_{\vec{p}\sigma} (\varepsilon_{\vec{p}}-\mu)} \nonumber \\
  &=& exp\left(-\frac{\varepsilon_{\vec{p}}-\mu}{T} \right),   \\ \label{9c}
  Z &=& \sum\limits_{\{n_{\vec{p}\sigma}\}}  \frac{1}{\prod\limits_{\vec{p}\sigma}n_{\vec{p}\sigma}!} \
  e^{-\frac{1}{T} \sum\limits_{\vec{p}\sigma} n_{\vec{p}\sigma} (\varepsilon_{\vec{p}}-\mu)} \nonumber \\
  &=& exp\left(\sum\limits_{\vec{p}\sigma} \langle n_{\vec{p}\sigma} \rangle  \right).
\end{eqnarray}
Substituting Eq.~(\ref{6c}) into Eq.~(\ref{1c}) and using Eq.~(\ref{9c}) and the Legendre transform (\ref{8b}), we obtain the entropy and the thermodynamic potential of the ideal gas as
\begin{eqnarray}\label{10c}
  S  &=&  -\sum\limits_{\vec{p}\sigma} [\langle n_{\vec{p}\sigma} \rangle \ln \langle n_{\vec{p}\sigma} \rangle -\langle n_{\vec{p}\sigma} \rangle],   \\ \label{11c}
 \Omega &=& T \sum\limits_{\vec{p}\sigma} \langle n_{\vec{p}\sigma} \rangle \left[\ln \langle n_{\vec{p}\sigma} \rangle -1 +\frac{\varepsilon_{\vec{p}}-\mu}{T}\right].
\end{eqnarray}
where $E=\sum_{\vec{p}\sigma} \langle n_{\vec{p}\sigma} \rangle \varepsilon_{\vec{p}}$ and $\langle N\rangle=\sum_{\vec{p}\sigma} \langle n_{\vec{p}\sigma} \rangle$.

In the Boltzmann-Gibbs statistics the single-particle distribution function (\ref{8c}) can also be found from the entropy of the ideal gas (\ref{10c}) by the extremization of the thermodynamic potential (\ref{11c}) with respect to $\langle n_{\vec{p}\sigma} \rangle$:
\begin{equation}\label{12c}
  \frac{\partial \Omega}{\partial \langle n_{\vec{p}\sigma} \rangle} = 0.
\end{equation}
Substituting Eq.~(\ref{11c}) into Eq.~(\ref{12c}), we obtain exactly the single-particle distribution function (\ref{8c}). Thus, in the Boltzmann-Gibbs statistics the single-particle distribution function $\langle n_{\vec{p}\sigma} \rangle$ derived from the entropy of the ideal gas (\ref{10c}) by the extremization of the thermodynamic potential (\ref{11c}) with respect to $\langle n_{\vec{p}\sigma} \rangle$ coincides with the distribution function $\langle n_{\vec{p}\sigma}\rangle$ of the Maxwell- Boltzmann ideal gas calculated from the general entropy (\ref{1c}) by the extremization of the thermodynamic potential (\ref{2c}) with respect to the probabilities of microstates of the system. Therefore, the constrained maximization of the entropy of the ideal gas with respect to the single-particle distribution function leads to the results of the Boltzmann-Gibbs statistics.

Let us show that this property of the Boltzmann-Gibbs statistics is not preserved for the Tsallis statistics if the Tsallis entropy of the ideal gas is generalized from the Boltzmann-Gibbs entropy of the ideal gas.

\subsection{Tsallis-$2$ statistics}
\subsubsection{The general formalism}
In the Tsallis-$2$ statistics the entropy and the thermodynamic potential of the grand canonical ensemble can be written as~\cite{Tsal98}
\begin{equation}\label{1d}
  S = -\sum\limits_{i} p_{i}^{q_{c}} \ln_{q_{c}} p_{i}
\end{equation}
and
\begin{equation}\label{2d}
  \Omega =  T\sum\limits_{i} p_{i}^{q_{c}} \left[\ln_{q_{c}} p_{i} + \frac{E_{i}-\mu N_{i}}{T}\right],
\end{equation}
where $\sum_{i} p_{i}=1$, $E=\sum_{i} p_{i}^{q_{c}} E_{i}$ and $\langle N\rangle=\sum_{i} p_{i}^{q_{c}} N_{i}$. The statistical averages of the Tsallis-$2$ statistics are defined as $\langle A \rangle=\sum_{i} p_{i}^{q_{c}} A_{i}$ (see ref.~\cite{Tsal98} and references therein). Note that Eq.~(\ref{1d}) is the same as Eq.~(\ref{1}) with the exception that here the parameter $q$ is denoted as $q_{c}$. Using the method of the Lagrange multipliers (\ref{3c})--(\ref{5c}) and Eq.~(\ref{2d}), we obtain
\begin{eqnarray}\label{3d}
p_{i} &=& \frac{1}{Z}  \left[1-(1-q_{c}) \frac{E_{i}-\mu N_{i}}{T} \right]^{\frac{1}{1-q_{c}}}, \\ \label{4d}
    Z &=& \sum\limits_{i} \left[1-(1-q_{c}) \frac{E_{i}-\mu N_{i}}{T} \right]^{\frac{1}{1-q_{c}}}.
\end{eqnarray}
In ref.~\cite{Tsal98}, these quantities were obtained only in the canonical ensemble.

Let us consider the Maxwell-Boltzmann ultrarelativistic ideal gas for $q_{c}>1$. Using Eq.~(\ref{11}) and the partition function (\ref{4d}), we obtain
\begin{eqnarray}\label{5d}
 Z &=& \sum\limits_{\{n_{\vec{p}\sigma}\}}  \frac{1}{\prod\limits_{\vec{p}\sigma}n_{\vec{p}\sigma}!} \
 \left[1-(1-q_{c}) \frac{\sum\limits_{\vec{p}\sigma} n_{\vec{p}\sigma} (\varepsilon_{\vec{p}}-\mu)}{T} \right]^{\frac{1}{1-q_{c}}} \nonumber \\
  &=&  \sum\limits_{N=0}^{N_{0}} \frac{\tilde{\omega}^{N}}{N!} a_{0}(0) \left[1+(1-q_{c}) \frac{\mu N}{T} \right]^{\frac{1}{1-q_{c}}+3N},
\end{eqnarray}
where
\begin{equation}\label{6d}
  a_{\eta}(\xi) = \frac{\Gamma\left(\frac{1}{q_{c}-1}+\xi-3(N+\eta)\right)}{(q_{c}-1)^{3(N+\eta)} \Gamma\left(\frac{1}{q_{c}-1}+\xi\right)}.
\end{equation}
The partition function (\ref{5d}) of the Tsallis-$2$ statistics is divergent as the norm equation (\ref{14}) of the Tsallis statistics. Therefore, the summation in Eq.~(\ref{5d}) was truncated by the upper bound $N_{0}$ to retain only the physical terms in the sum. To find the upper bound of summation $N_{0}$, we rewrite the partition function (\ref{5d}) in the form $Z=\sum_{N} \phi(N)$, where
\begin{equation}\label{6da}
  \phi(N) =  \frac{\tilde{\omega}^{N}}{N!} a_{0}(0) \left[1+(1-q_{c}) \frac{\mu N}{T} \right]^{\frac{1}{1-q_{c}}+3N}.
\end{equation}
Then the cut-off parameter $N_{0}$ can be found either from the inflection point of the function $\ln\phi(N)$ substituting Eq.~(\ref{6da}) into Eq.~(\ref{17b}) or from the local minimum of this function $\ln\phi(N)$.

The mean occupation numbers for the Maxwell- Boltzmann ultrarelativistic ideal gas in the Tsallis-$2$ statistics in the grand canonical ensemble for $q_{c}>1$ can be written as
\begin{eqnarray}\label{7d}
  \langle n_{\vec{p}\sigma}\rangle  &=& \frac{1}{Z^{q_{c}}}\sum\limits_{\{n_{\vec{p}\sigma}\}}  n_{\vec{p}\sigma}
  \frac{1}{\prod\limits_{\vec{p}\sigma}n_{\vec{p}\sigma}!} \nonumber \\
  && \left[1-(1-q_{c}) \frac{\sum\limits_{\vec{p}\sigma} n_{\vec{p}\sigma} (\varepsilon_{\vec{p}}-\mu)}{T} \right]^{\frac{q_{c}}{1-q_{c}}} \nonumber \\
  &=& \frac{1}{Z^{q_{c}}} \sum\limits_{N=0}^{N_{0}} \frac{\tilde{\omega}^{N}}{N!} a_{0}(1) \nonumber \\ && \left[1+(q_{c}-1) \frac{\varepsilon_{\vec{p}}-\mu(N+1)}{T} \right]^{\frac{q_{c}}{1-q_{c}}+3N},
\end{eqnarray}
where the upper bound of summation $N_{0}$ is the same as in Eq.~(\ref{5d}) and $a_{0}(1)$ is calculated from Eq.~(\ref{6d}).

The mean number of particles for the Maxwell- Boltzmann ultrarelativistic ideal gas in the Tsallis-$2$ statistics for $q_{c}>1$ takes the form
\begin{eqnarray}\label{9d}
 \langle N \rangle &=& \frac{1}{Z^{q_{c}}}  \sum\limits_{N=0}^{N_{0}} \frac{\tilde{\omega}^{N+1}}{N!} a_{1}(1) \nonumber \\ && \left[1+(1-q_{c}) \frac{\mu (N+1)}{T} \right]^{\frac{q_{c}}{1-q_{c}}+3(N+1)},
\end{eqnarray}
where the upper bound of summation $N_{0}$ is the same as in Eq.~(\ref{5d}) and $a_{1}(1)$ is calculated from Eq.~(\ref{6d}). The energy for the Maxwell- Boltzmann ultrarelativistic ideal gas in the Tsallis-$2$ statistics in the grand canonical ensemble for $q_{c}>1$ is given by
\begin{eqnarray}\label{11d}
 E &=& \frac{3T}{Z^{q_{c}}}  \sum\limits_{N=0}^{N_{0}} \frac{\tilde{\omega}^{N+1}}{N!} a_{1}(0) \nonumber \\ && \left[1+(1-q_{c}) \frac{\mu (N+1)}{T} \right]^{\frac{1}{1-q_{c}}+3(N+1)},
\end{eqnarray}
where the upper bound of summation $N_{0}$ is the same as in Eq.~(\ref{5d}) and $a_{1}(0)$ is calculated from Eq.~(\ref{6d}). Then the entropy (\ref{1d}) for the Maxwell- Boltzmann ultrarelativistic ideal gas in the Tsallis-$2$ statistics in the grand canonical ensemble for $q_{c}>1$ can be rewritten as
\begin{equation}\label{13d}
  S= \frac{1}{Z^{1-q_{c}}} \left[Q \ln_{q_{c}} Z + \frac{E-\mu \langle N \rangle }{T} \right],
\end{equation}
where
\begin{eqnarray}\label{14d}
 Q &\equiv& \sum\limits_{i} p_{i}^{q_{c}} \nonumber \\
  &=& \frac{1}{Z^{q_{c}}} \sum\limits_{N=0}^{N_{0}} \frac{\tilde{\omega}^{N}}{N!} a_{0}(1) \left[1+(1-q_{c}) \frac{\mu N}{T} \right]^{\frac{q_{c}}{1-q_{c}}+3N}. \;\;\;\;\;
\end{eqnarray}
Here the upper bound of summation $N_{0}$ is the same as in Eq.~(\ref{5d}) and $a_{0}(1)$ is calculated from Eq.~(\ref{6d}). Finally, the thermodynamic potential (\ref{2d}) for the Maxwell- Boltzmann ultrarelativistic ideal gas in the Tsallis-$2$ statistics in the grand canonical ensemble for $q_{c}>1$ is
\begin{eqnarray}\label{14da}
 \Omega &=& -\tau \ln_{q_{c}} Z, \\ \label{14db}
 \tau &=& \frac{T}{Z^{1-q_{c}}} \left[Q - (1-q_{c}) \frac{E-\mu \langle N \rangle }{T} \right].
\end{eqnarray}
Note that in the Gibbs limit $q_{c}\to 1$ the thermodynamic quantities (\ref{5d}), (\ref{7d}), (\ref{9d}), (\ref{11d}), (\ref{13d}) and (\ref{14da}) resemble their Boltzmann- Gibbs values. The energy (\ref{11d}) of the Tsallis-$2$ statistics leads to the constraint that $q_{c}<4/3$.

The ultrarelativistic transverse momentum distribution of the Tsallis-$2$ statistics can be written as
\begin{eqnarray}\label{14dc}
  \frac{d^{2}N}{dp_{T}dy} &=& \frac{gV}{(2\pi)^{2}} p_{T}^{2} \cosh y  \
 \frac{1}{Z^{q_{c}}} \sum\limits_{N=0}^{N_{0}} \frac{\tilde{\omega}^{N}}{N!} a_{0}(1) \nonumber \\ && \left[1+(q_{c}-1) \frac{p_{T} \cosh y-\mu(N+1)}{T} \right]^{\frac{q_{c}}{1-q_{c}}+3N}, \;\;\;\;\;\;
\end{eqnarray}
where the upper bound of summation $N_{0}$ is the same as in Eq.~(\ref{5d}) and $a_{0}(1)$ is calculated from Eq.~(\ref{6d}). In the Gibbs limit $q_{c}\to 1$ Eq.~(\ref{14dc}) recovers the Maxwell - Boltzmann transverse momentum distribution of the Boltzmann - Gibbs statistics (\ref{24a}).

\subsubsection{The zeroth term approximation of the Tsallis-$2$ statistics}
Let us rewrite the thermodynamic quantities of the Tsallis-$2$ statistics in the case of the zeroth term approximation taking into the summations only the terms with $N=0$. Then the partition function (\ref{5d}) is equal to unity, $Z=1$. The mean occupation numbers (\ref{7d}) in the zeroth term approximation can be written as
\begin{equation}\label{15d}
  \langle n_{\vec{p}\sigma}\rangle = \left[1+(q_{c}-1)\frac{\varepsilon_{\vec{p}}-\mu}{T}\right]^{\frac{q_{c}}{1-q_{c}}}.
\end{equation}
The Tsallis-$2$ mean number of particles (\ref{9d}) for $N=0$ takes the form
\begin{equation}\label{16d}
  \langle N \rangle = \frac{gVT^{3}}{\pi^{2}}
  \frac{\Gamma\left(\frac{q_{c}}{q_{c}-1}-3\right)}{\Gamma\left(\frac{q_{c}}{q_{c}-1}\right) (q_{c}-1)^{3}} \left[1-(q_{c}-1)\frac{\mu}{T}\right]^{\frac{q_{c}}{1-q_{c}}+3}.
\end{equation}
The Tsallis-$2$ energy (\ref{11d}) in the zeroth term approximation is given by
\begin{equation}\label{17d}
  E = \frac{3gVT^{4}}{\pi^{2}}
  \frac{\Gamma\left(\frac{q_{c}}{q_{c}-1}-4\right)}{\Gamma\left(\frac{q_{c}}{q_{c}-1}\right) (q_{c}-1)^{4}} \left[1-(q_{c}-1)\frac{\mu}{T}\right]^{\frac{q_{c}}{1-q_{c}}+4}.
\end{equation}
In the zeroth term approximation of the Tsallis-$2$ statistics we have the condition that $q_{c}<4/3$.

The entropy (\ref{13d}) in the zeroth term approximation can be rewritten as
\begin{eqnarray}\label{18d}
  S &=& \left(3-\frac{\mu}{T}\right) \frac{gVT^{3}}{\pi^{2}}
  \frac{\Gamma\left(\frac{q_{c}}{q_{c}-1}-4\right)}{\Gamma\left(\frac{q_{c}}{q_{c}-1}\right) (q_{c}-1)^{4}} \nonumber \\
  && \left[1-(q_{c}-1)\frac{\mu}{T}\right]^{\frac{q_{c}}{1-q_{c}}+3}.
\end{eqnarray}
Using Eqs.~(\ref{15d}) and (\ref{18d}), we obtain
\begin{equation}\label{19d}
  S=-\sum\limits_{\vec{p}\sigma} \langle n_{\vec{p}\sigma}\rangle \ln_{q_{c}}\langle n_{\vec{p}\sigma}\rangle^{1/q_{c}}.
\end{equation}
The thermodynamic potential (\ref{14da}) in the zeroth term approximation is equal to zero, $\Omega=0$ and $\tau=T [1-(1-q_{c})S]$. Note that in the Gibbs limit $q_{c}\to 1$ the mean occupation numbers (\ref{15d}), the mean number of particles (\ref{16d}) and the energy (\ref{17d}) recover their Boltzmann-Gibbs values. However, the thermodynamic potential $\Omega$ and the entropy $S$ in the zeroth term approximation do not resemble their corresponding relations of the Boltzmann-Gibbs statistics. Hence, the zeroth term approximation of the Tsallis-$2$ statistics is not consistent.

Let us introduce the transformation $q_{c}\to 1/q$. Then Eqs.~(\ref{15d}), (\ref{16d}) and (\ref{17d}) exactly coincide with Eqs.~(\ref{26}), (\ref{27}) and (\ref{27a}), respectively, of the Tsallis statistics in the zeroth term approximation. However, under this substitution the entropy (\ref{19d}) becomes
\begin{equation}\label{20d}
  S=-q\sum\limits_{\vec{p}\sigma} \langle n_{\vec{p}\sigma}\rangle^{q} \ln_{q}\langle n_{\vec{p}\sigma}\rangle.
\end{equation}
Equation (\ref{20d}) is not equal to Eq.~(\ref{27d}) of the Tsallis statistics in the zeroth term approximation. Moreover, the thermodynamic potential $\Omega$ under this transformation does not also recover the thermodynamic potential (\ref{27b}) of the Tsallis statistics in the zeroth term approximation.

The ultrarelativistic transverse momentum distribution of the Tsallis-$2$ statistics (\ref{14dc}) in the zeroth term approximation can be rewritten as
\begin{equation}\label{21d}
  \frac{d^{2}N}{dp_{T}dy} = \frac{gV p_{T}^{2} \cosh y}{(2\pi)^{2}} \left[1+(q_{c}-1) \frac{p_{T} \cosh y-\mu}{T} \right]^{\frac{q_{c}}{1-q_{c}}}.
\end{equation}
Thus, the ultrarelativistic transverse momentum distribution (\ref{21d}) of the Tsallis-$2$ statistics in the zeroth term approximation exactly coincides with the ultrarelativistic transverse momentum distribution (\ref{30}) of the Tsallis-factorized statistics.

\subsection{Tsallis-factorized statistics}
The Tsallis-factorized statistics is defined by the generalized entropy of the ideal gas, the generalized mean number of particles and the generalized energy, which in the case of the Maxwell- Boltzmann ideal gas have the form~\cite{Cleymans12a,Cleymans2012}
\begin{equation}\label{1z}
   S=-\sum\limits_{\vec{p}\sigma} \left[f_{\vec{p}\sigma}^{q_{c}} \ln_{q_{c}} f_{\vec{p}\sigma} - f_{\vec{p}\sigma} \right]
\end{equation}
and
\begin{eqnarray}\label{2z}
  \langle N \rangle &=& \sum\limits_{\vec{p}\sigma} f_{\vec{p}\sigma}^{q_{c}},  \\ \label{3z}
  E &=& \sum\limits_{\vec{p}\sigma} f_{\vec{p}\sigma}^{q_{c}} \varepsilon_{\vec{p}},
\end{eqnarray}
where $f_{\vec{p}\sigma}$ is the single-particle distribution function and $f_{\vec{p}\sigma}^{q_{c}}\equiv\langle n_{\vec{p}\sigma}\rangle$. The thermodynamic potential $(\Omega=E-TS-\mu \langle N \rangle)$ of the ideal gas in the Tsallis-factorized statistics in the grand canonical ensemble can be written as
\begin{equation}\label{4z}
   \Omega = T\sum\limits_{\vec{p}\sigma} f_{\vec{p}\sigma}^{q_{c}} \left[q_{c} \ln_{q_{c}} f_{\vec{p}\sigma} - 1 + \frac{\varepsilon_{\vec{p}}-\mu}{T} \right].
\end{equation}
The unknown single-particle distribution function $f_{\vec{p}\sigma}$ is obtained from the local extrema of the thermodynamic potential (\ref{4z}):
\begin{equation}\label{5z}
  \frac{\partial \Omega}{\partial f_{\vec{p}\sigma}} = 0.
\end{equation}
Then the single-particle distribution function of the Tsallis-factorized statistics in the grand canonical ensemble becomes~\cite{Cleymans12a,Cleymans2012}
\begin{equation}\label{6z}
 f_{\vec{p}\sigma}^{q_{c}} = \langle n_{\vec{p}\sigma}\rangle = \left[1+(q_{c}-1)\frac{\varepsilon_{\vec{p}}-\mu}{T}\right]^{\frac{q_{c}}{1-q_{c}}}.
\end{equation}
The mean occupation numbers (\ref{6z}) of the Tsallis-factorized statistics are equivalent to the mean occupation numbers (\ref{15d}) of the Tsallis-$2$ statistics in the zeroth term approximation. However, they are not equivalent to the exact mean occupation numbers (\ref{7d}) of the Tsallis-$2$ statistics.

Let us consider the ultrarelativistic ideal gas for $q_{c}>1$. Substituting Eq.~(\ref{6z}) into Eqs.~(\ref{2z}) and (\ref{3z}), we obtain the mean number of particles (\ref{16d}) and the energy (\ref{17d}) of the Tsallis-$2$ statistics in the zeroth term approximation. Substituting Eq.~(\ref{6z}) into Eqs.~(\ref{1z}) and (\ref{4z}), and using Eqs.~(\ref{16d}) and (\ref{17d}), we obtain
\begin{eqnarray}\label{7z}
  S &=& \left(4-q_{c}\frac{\mu}{T}\right) \frac{gVT^{3}}{\pi^{2}}
  \frac{\Gamma\left(\frac{q_{c}}{q_{c}-1}-4\right)}{\Gamma\left(\frac{q_{c}}{q_{c}-1}\right) (q_{c}-1)^{4}} \nonumber \\
  && \left[1-(q_{c}-1)\frac{\mu}{T}\right]^{\frac{q_{c}}{1-q_{c}}+3}
\end{eqnarray}
and
\begin{equation}\label{8z}
  \Omega = -T \frac{gVT^{3}}{\pi^{2}}
  \frac{\Gamma\left(\frac{q_{c}}{q_{c}-1}-4\right)}{\Gamma\left(\frac{q_{c}}{q_{c}-1}\right) (q_{c}-1)^{4}} \left[1-(q_{c}-1)\frac{\mu}{T}\right]^{\frac{q_{c}}{1-q_{c}}+4}.
\end{equation}
The entropy (\ref{7z}) and the thermodynamic potential (\ref{8z}) differ from the corresponding entropy (\ref{18d}) and thermodynamic potential $\Omega=0$ of the ideal gas of the Tsallis-$2$ statistics in the zeroth term approximation. In the Gibbs limit $q_{c}\to 1$ the thermodynamic quantities (\ref{1z})--(\ref{4z}) and (\ref{6z})--(\ref{8z}) resemble their Boltzmann-Gibbs values. The condition that the entropy (\ref{7z}) has a finite value leads to the constraint that $q_{c}<4/3$ as in the case of the Tsallis-$2$ statistics.

The mean occupation numbers (\ref{6z}) of the Tsallis- factorized statistics derived from the generalized entropy of the ideal gas (\ref{1z}) by the extremization of the thermodynamic potential (\ref{4z}) with respect to $f_{\vec{p}\sigma}$ are not equivalent to the mean occupation numbers (\ref{7d}) of the ideal gas of the Tsallis-$2$ statistics calculated from the constrained maximization of the Tsallis entropy (\ref{1d}) with respect to the probabilities of microstates of the system. This means that the Tsallis-factorized statistics defined on the basis of the generalized entropy of the ideal gas (\ref{1z}) is not equivalent to the Tsallis-$2$ statistics. By definition, the Tsallis-factorized statistics is based  on the generalized property of the ideal gas of the Boltzmann-Gibbs statistics which states that the single-particle distribution function obtained from the constrained maximization of the Boltzmann- Gibbs entropy of the ideal gas with respect to the probing single-particle distribution function is the same as the single-particle distribution function of the Boltzmann- Gibbs statistics itself. Therefore, the constrained maximization of the Tsallis-factorized entropy of the ideal gas generalized from the Boltzmann-Gibbs entropy of the ideal gas with respect to the single-particle distribution function should lead to the results of the Tsallis-$2$ statistics. However, they are differ. Therefore, the Tsallis-factorized statistics is not equivalent to the Tsallis statistics (the Tsallis-$2$ statistics).

Under the substitution $q_{c}\to 1/q$ the mean occupation numbers (\ref{6z}) of the Tsallis-factorized statistics are transformed into the mean occupation numbers (\ref{26}) of the Tsallis statistics in the zeroth term approximation. However, the generalized entropy of the ideal gas (\ref{1z}) under this substitution does not recover the entropy (\ref{27d}) of the ideal gas in the zeroth term approximation. Moreover, under this substitution all the thermodynamic quantities of the Tsallis-factorized statistics do not recover their corresponding exact relations of the Tsallis statistics. Thus, the Tsallis-factorized statistics defined on the basis of the generalized entropy of the ideal gas (\ref{1z}) is not equivalent to the Tsallis statistics either.

It should be emphasised that both the Tsallis-$2$ statistics and the Tsallis-factorized statistics are defined on the basis of the generalized expectation values. Nevertheless, the ultrarelativistic transverse momentum distribution of the Tsallis-factorized statistics (\ref{30}) does not recover the ultrarelativistic transverse momentum distribution (\ref{14dc}) of the Tsallis-$2$ statistics because the transverse momentum distribution of the Tsallis-factorized statistics corresponds only to the zero term ($N=0$) in the sum (\ref{14dc}) for the transverse momentum distribution of the Tsallis-$2$ statistics. The presence of the higher-order terms ($N\geq 1$) in the sum (\ref{14dc}) for the transverse momentum distribution of the Tsallis-$2$ statistics is related to the fact that the Tsallis-$2$ statistics is defined on the basis of the probabilities of microstates of the system. However, the absence of such terms in the transverse momentum distribution of the Tsallis-factorized statistics (\ref{30}) is explained by the fact that the Tsallis-factorized statistics is defined on the basis of the single-particle distribution functions of the ideal gas, which correspond to the zeroth term approximation of the Tsallis-$2$ statistics at $N_{0}=0$. The equivalence of the transverse momentum distribution (\ref{21d}) of the Tsallis-$2$ statistics in the zeroth term approximation with the transverse momentum distribution (\ref{30}) of the Tsallis-factorized statistics can be explained by the fact that the Tsallis-factorized statistics is defined on the basis of the single-particle distribution functions of the ideal gas and the zeroth term approximation of the Tsallis-$2$ statistics is defined on the basis of the quantities in which the higher-order terms ($N\geq 1$) were neglected.

Hence, the Tsallis-factorized statistics does not correspond to the Tsallis-$2$ statistics due to the different definitions of the probability of states.

\section{Discussion and conclusions}\label{sec5}
In the present paper, the analytical results for the transverse momentum distribution and some thermodynamic quantities of the ultrarelativistic ideal gas of the Tsallis statistics in the grand canonical ensemble were obtained. The thermodynamic quantities were defined in the general form in terms of the probabilities of microstates. After that these thermodynamic quantities and momentum distribution were rewritten in the occupation number representation and were solved using the integral representation of the Gamma-function. The thermodynamic quantities were expanded in the power series indexed by the natural numbers $N$. It was revealed that such infinite sums over index $N$ are divergent for $q<1$ because the power-law function in comparison with the exponent cannot suppress the rapid growth of the number of microstates of the system with increasing energy and number of particles of the system. To suppress them, we introduced the regularization scheme based on the cut-off of such infinite sums. For $q>1$ we use the Tsallis cut-off prescription.

In the case of $q<1$, we have introduced the zeroth term approximation by imposing the upper bound of summation $N_{0}=0$. This approximation is reliable for the Tsallis statistics only at large deviations of the parameter $q$ from unity; however, it is generally inconsistent. The analytical formulas for the transverse momentum distribution and some thermodynamic quantities in the zeroth term approximation were found. We have revealed that the transverse momentum distribution of the Tsallis-factorized statistics, which is widely used to describe the experimental momentum distributions of hadrons at high energies, in the case of massless particles corresponds to the transverse momentum distribution of the Tsallis statistics in the zeroth term approximation with transformation of the parameter $q$ to $1/q_{c}$.

In this paper, the Tsallis-$2$ statistics and the Tsallis-factorized statistics were also considered in the general form. We have analytically demonstrated on the basis of the ultrarelativistic ideal gas that the Tsallis-factorized statistics is not equivalent to the Tsallis and the Tsallis-$2$ statistics. The constrained maximization of the Tsallis-factorized entropy of the ideal gas with respect to the single-particle distribution function does not lead to the true results for the Tsallis and the Tsallis-$2$ statistics. In particular, the Tsallis-factorized distribution is not equivalent to the distributions of the Tsallis and the Tsallis-$2$ statistics. However, the mean occupation numbers of the Tsallis-factorized statistics are similar to the mean occupation numbers of the Tsallis-$2$ statistics in the zeroth term approximation. Thus, the Tsallis- factorized statistics may be an additional particular statistics independent of the Tsallis and the Tsallis-$2$ statistics.

\begin{acknowledgement}
This work was supported in part by the joint research project and grant of JINR and IFIN-HH (protocol N~4543). I am indebted to D.-V.~Anghel, J.~Cleymans, G.I.~Lykasov, A.S.~Sorin and O.V.~Teryaev for stimulating discussions.
\end{acknowledgement}

\appendix

\section{Regularization of the norm equation}\label{App1}
The norm equation (\ref{5}) of the Tsallis statistics for the ultrarelativistic Maxwell-Boltzmann particles in the grand canonical ensemble can be rewritten as
\begin{equation}\label{a1}
  \sum\limits_{N=0}^{\infty} \int\limits_{0}^{\infty} dE  \left[1+\frac{1}{z+1}\frac{\Lambda-E+\mu N}{T}\right]^{z} W_{N,E} = 1,
\end{equation}
where
\begin{equation}\label{a2}
  W_{N,E}=\sum\limits_{\{n_{\vec{p}\sigma}\}} \frac{1}{\prod\limits_{\vec{p}\sigma}n_{\vec{p}\sigma}!} \ \delta(\sum\limits_{\vec{p}\sigma} n_{\vec{p}\sigma}-N)
  \delta(\sum\limits_{\vec{p}\sigma} n_{\vec{p}\sigma} \varepsilon_{\vec{p}} - E).
\end{equation}
Here, $N$ and $E$ are the number of particles and energy of system, respectively. The weight (\ref{a2}) for the Maxwell-Boltzmann ultrarelativistic ideal gas in the microcanonical ensemble takes the form
\begin{equation}\label{a3}
   W_{N,E}=\frac{1}{N!} \left(\frac{gV}{\pi^{2}}\right)^{N} \frac{E^{3N-1}}{\Gamma(3N)}.
\end{equation}
Substituting Eq.~(\ref{a3}) into Eq.~(\ref{a1}) we obtain
\begin{eqnarray}\label{a4}
  \sum\limits_{N=0}^{\infty} && \frac{1}{N!} \left(\frac{gV}{\pi^{2}}\right)^{N} \frac{1}{\Gamma(3N)} \nonumber \\
  && \int\limits_{0}^{\infty} dE  \left[1+\frac{1}{z+1}\frac{\Lambda-E+\mu N}{T}\right]^{z} E^{3N-1} = 1. \;\;\;
\end{eqnarray}
For negative values of $z$ $(q<1)$ at large $E\to\infty$ the integral function in Eq.~(\ref{a4}) is proportional to $\sim E^{3N-1+z}$. Thus, the integral is convergent only if $N<-z/3$. This condition implies the cut-off on the sum over $N$. In the case of positive values of $z$ $(q>1)$, at large $E\to\infty$ the argument of the power-law function can be negative. Therefore, if $z$ is fixed, the Tsallis cut-off $\theta(1+(\Lambda-E+\mu N)/((z+1)T))$, where $\theta(x)$ is the step function, should be imposed.

Integrating Eq.~(\ref{a4}), we obtain
\begin{equation}\label{a5}
  \sum\limits_{N=0}^{\infty} \frac{1}{N!} \left(\frac{gVT^{3}}{\pi^{2}}\right)^{N} h_{0}(0) \left[1+\frac{\Lambda+\mu N}{(z+1)T}\right]^{z+3N}=1,
\end{equation}
where
\begin{eqnarray}\label{a6}
   h_{0}(0) &=& \frac{(-z-1)^{3N}\Gamma(-z-3N)}{\Gamma(-z)}, \quad  z<-1, \\ \label{a7}
   h_{0}(0) &=& \frac{(z+1)^{3N}\Gamma(z+1)}{\Gamma(z+1+3N)}, \qquad \qquad z>0.
\end{eqnarray}
The sum over $N$ in Eq.~(\ref{a5}) is constrained by the conditions $-z-3N>0$ and $1+(\Lambda+\mu N)/((z+1)T)>0$ for $z<-1$, and the conditions $z+1+3N<\infty$ and $1+(\Lambda+\mu N)/((z+1)T)>0$ for $z>0$. These constraints truncate the sum in Eq.~(\ref{a5}) and regularize it allowing to obtain the physical results.


\begin{thebibliography}{}
%

\bibitem{ALICE09} K.~Aamodt et al. (ALICE Collaboration), Eur. Phys. J. C \textbf{71}, 1655 (2011).

\bibitem{Atlas1} G.~Aad et al. (ATLAS Collaboration), New J. Phys. \textbf{13}, 053033 (2011).

\bibitem{Cms3} V.~Khachatryan et al. (CMS Collaboration), Phys. Rev. Lett. \textbf{105}, 022002 (2010).


\bibitem{Rybczynski14} M.~Rybczy\'{n}ski, Z.~W{\l}odarczyk, Eur. Phys. J. C \textbf{74}, 2785 (2014).

\bibitem{Cleymans13} J.~Cleymans, G.I.~Lykasov, A.S.~Parvan, A.S.~Sorin, O.V.~Teryaev, D.~Worku, Phys. Lett. B \textbf{723}, 351 (2013).

\bibitem{Azmi14} M.D.~Azmi, J.~Cleymans, J. Phys. G: Nucl. Part. Phys. \textbf{41}, 065001 (2014).

\bibitem{Cleymans12a} J.~Cleymans, D.~Worku, J. Phys. G: Nucl. Part. Phys. \textbf{39}, 025006 (2012).

\bibitem{Cleymans2012}  J.~Cleymans, D.~Worku, Eur. Phys. J. A \textbf{48}, 160 (2012).

\bibitem{Marques13} L.~Marques, E.~Andrade-II, A.~Deppman, Phys. Rev. D \textbf{87}, 114022 (2013).

\bibitem{Li14} B.-C.~Li, Y.-Z.~Wang, F.-H.~Liu, X.-J.~Wen and Y.-E.~Dong, Phys. Rev. D \textbf{89}, 054014 (2014).

\bibitem{Parvan14} A.S.~Parvan, PoS (Baldin-ISHEPP-XXII), 077 (2015).

\bibitem{Parvan16a} A.S.~Parvan, O.V.~Teryaev, J.~Cleymans,	arXiv: 1607. 01956v2 [nucl-th].

\bibitem{Biyajima06} M.~Biyajima, T.~Mizoguchi, N.~Nakajima, N.~Suzuki, G.~Wilk, Eur. Phys. J. C \textbf{48}, 597 (2006).

\bibitem{Marques15} L.~Marques, J.~Cleymans, A.~Deppman, Phys. Rev. D \textbf{91}, 054025 (2015).


\bibitem{Tsal88} C.~Tsallis, J. Stat. Phys. \textbf{52}, 479 (1988).

\bibitem{Tsal98} C.~Tsallis, R.S.~Mendes, A.R.~Plastino, Physica A \textbf{261}, 534 (1998).


\bibitem{Tirnakli00} U.~Tirnakli, D.F.~Torres, Eur. Phys. J. B \textbf{14}, 691 (2000).


\bibitem{Buyukkilic93} F.~B\"{u}y\"{u}kkili\c{c}, D.~Demirhan, A.~G\"{u}le\c{c}, Phys. Lett. A \textbf{197}, 209 (1995).


\bibitem{Lavagno02} A.~Lavagno, Phys. Lett. A \textbf{301}, 13 (2002).

\bibitem{Alberico09} W.M.~Alberico, A.~Lavagno, Eur. Phys. J. A \textbf{40}, 313 (2009).

\bibitem{Conroy10} J.M.~Conroy, H.G.~Miller, A.R.~Plastino, Phys. Lett. A \textbf{374}, 4581 (2010).


\bibitem{Parvan2015} A.S.~Parvan, Eur. Phys. J. A \textbf{51}, 108 (2015).

\bibitem{Parvan06a} A.S.~Parvan, Phys. Lett. A \textbf{350}, 331 (2006).

\bibitem{Parvan06b} A.S.~Parvan, Phys. Lett. A \textbf{360}, 26 (2006).



\bibitem{Parvan2015a} A.S.~Parvan, Foundation of equilibrium statistical mechanics based on generalized entropy, in {\it Recent Advances in Thermo and Fluid Dynamics}, ed. by Mofid Gorji-Bandpy (InTech, Rijeka, 2015), p.303.


\bibitem{TsallCutOff} A.M.~Teweldeberhan, A.R.~Plastino, H.G.~Miller, Phys. Lett. A \textbf{343}, 71 (2005).

\bibitem{Abramowitz} M.~Abramowitz, I.~Stegun, \textit{Handbook of Mathematics Functions, Nat. Bur. Stand. Appl. Math. Ser., vol. 55} (U.S. Govt. Printing Office, Washington, DC 1965).

\bibitem{Prato} D.~Prato, Phys. Lett. A \textbf{203}, 165 (1995).


\bibitem{Parvan16} A.S.~Parvan, Eur. Phys. J. A \textbf{52}, 355 (2016).


\end{thebibliography}
\end{document}